\documentclass[epj,twocolumn]{webofc}
\usepackage[varg]{txfonts}   
\usepackage{wrapfig}
\usepackage{braket}
\def\JP{J\!/\psi}
\woctitle{Flavour changing and conserving processes}
\begin{document}
\title{Pseudoscalar transition form factors: $(g-2)$ of the muon, pseudoscalar decays into lepton pairs, and the $\eta-\eta'$ mixing}
%
%

\author{Pere Masjuan\inst{1}\fnsep\thanks{\email{masjuan@kph.uni-mainz.de}} \and
        Pablo Sanchez-Puertas\inst{1}\fnsep\thanks{\email{sanchezp@kph.uni-mainz.de}}         
}

\institute{PRISMA Cluster of Excellence, Institut f\"ur Kernphysik, Johannes Gutenberg-Universit\"at, D-55099 Mainz, Germany}

\abstract{%
We present our model-independent and data-driven method to describe pseudoscalar meson transition form factors in the space- and (low-energy) time-like regions. The method is general and conforms a toolkit applicable to any other form factor, of one and two variables, with the potential to include both high- and low-energy QCD constraints altogether. The method makes use of analyticity and unitary properties of form factors, it is simple, systematic and can be improved upon by including new data. In the present discussion, the method is used to show the impact of experimental data for precision calculations in the low-energy sector of the Standard Model. In particular, due to its relevance for New Physics searches, we have considered the hadronic light-by-light scattering contribution to the anomalous magnetic moment of the muon (the pseudoscalar exchange contribution), the pseudoscalar decays into lepton pairs, and the determination of the mixing parameters of the $\eta$ and $\eta'$ system. For all of them we provide the most updated results in a data-driven manner.
}
%
%
\onecolumn
\thispagestyle{empty}

\vfill

\begin{center}
{\huge\bf
Pseudoscalar transition form factors: \\[2mm]
$(g-2)$ of the muon, pseudoscalar decays into lepton pairs, and the $\eta-\eta'$ mixing}\\[1.5cm]
{\large\bf Pere Masjuan$^*$ and Pablo Sanchez-Puertas}\\[1cm]
{\large PRISMA Cluster of Excellence, Institut f\"ur Kernphysik, Johannes Gutenberg-Universit\"at,\\[1mm]
 D-55099 Mainz, Germany}

\vfill

{\large\bf Abstract}\\[3mm]

\begin{minipage}{0.8\textwidth}
We present our model-independent and data-driven method to describe pseudoscalar meson transition form factors in the space- and (low-energy) time-like regions. The method is general and conforms a toolkit applicable to any other form factor, of one and two variables, with the potential to include both high- and low-energy QCD constraints altogether. The method makes use of analyticity and unitary properties of form factors, it is simple, systematic and can be improved upon by including new data. In the present discussion, the method is used to show the impact of experimental data for precision calculations in the low-energy sector of the Standard Model. In particular, due to its relevance for New Physics searches, we have considered the hadronic light-by-light scattering contribution to the anomalous magnetic moment of the muon (the pseudoscalar exchange contribution), the pseudoscalar decays into lepton pairs, and the determination of the mixing parameters of the $\eta$ and $\eta'$ system. For all of them we provide the most updated results in a data-driven manner.
\end{minipage}
\end{center}
\vfill
\noindent\rule{8cm}{0.5pt}\\
$^*$ Invited talk at the FCCP2015 - Workshop on ``Flavour changing and conserving processes,'' 10-12 September 2015,Anacapri, Capri Island, Italy.
\setcounter{page}{0}
\newpage
\twocolumn
\maketitle
\section{Introduction}
\label{intro}

The pseudoscalar meson transition form factors (TFF) describe the effect of the strong interaction on the $\gamma^*\gamma^* - P$ transition (where $P=\pi^0, \eta, \eta^\prime\cdots$) and is represented by a function $F_{P\gamma^*\gamma^*}(q_1^2,q_2^2)$ of the photon virtualities $q_1^2$, and $q_2^2$.

\begin{figure}
\centering
\includegraphics[width=10.0cm,trim=10cm 12cm 5.5cm 5cm, clip=true]{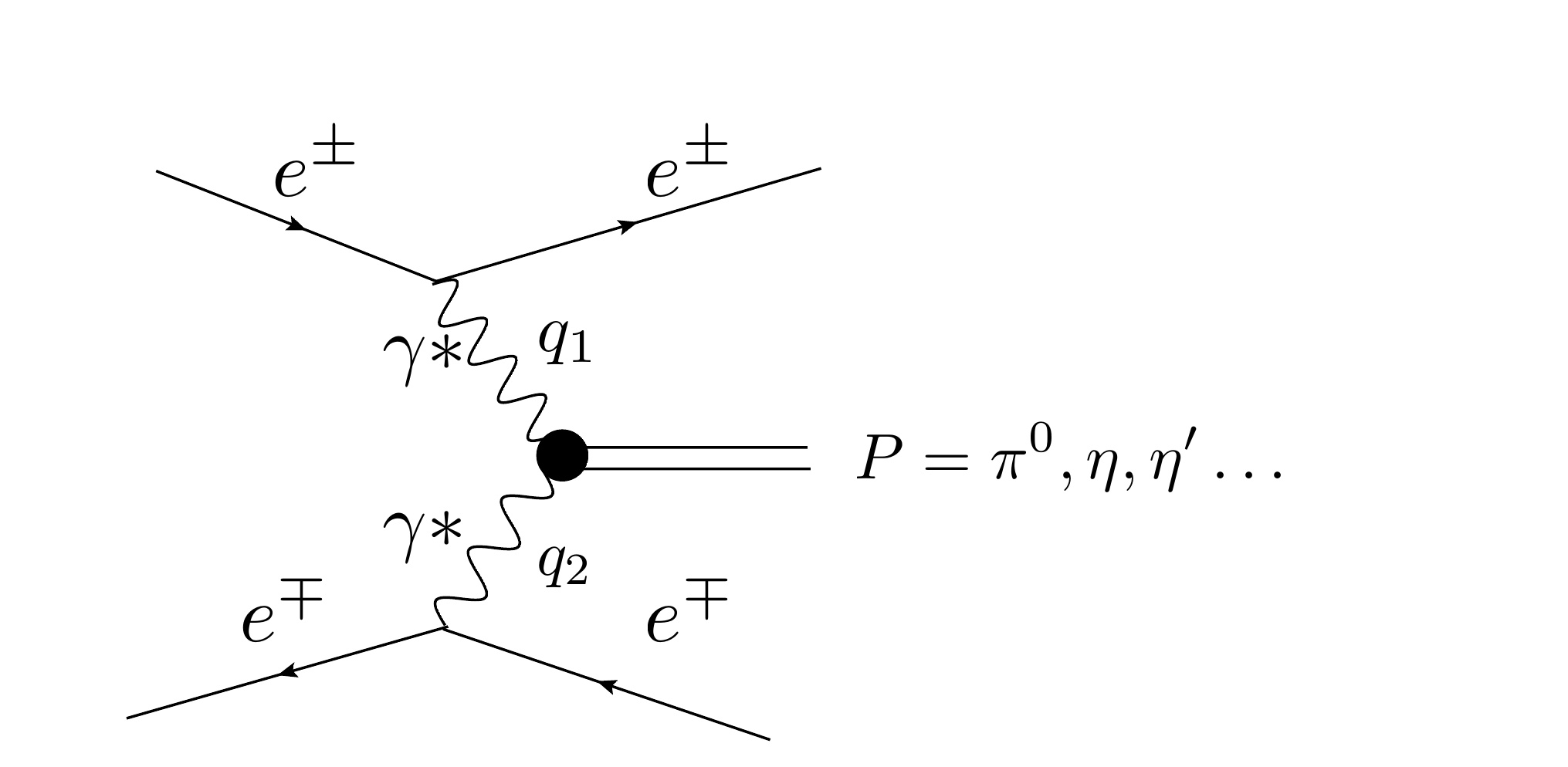}
\caption{$e^+e^- \to e^+e^- P$ space-like process. $q_1$ and $q_2$ represent the photon virtualities and the black blob in the interaction point represents the meson transition form factor $F_{P\gamma^*\gamma^*}(q_1^2,q_2^2)$.}
\label{fig:SL}       
\end{figure}

From the experimental point of view, one can study the TFF from both space- and time-like energy regimes. The time-like TFF can be accessed from a single Dalitz decay $P \to l^+l^- \gamma$ process which contains an off-shell photon with the momentum transfer $q_1^2$, a $F_{P\gamma^*\gamma}(q_1^2,0)$, covering the $4m_l^2<q^2<m_P^2$ region.  The space-like TFF can be accessed in $e^+e^-$ colliders by the two-photon-fusion reaction $e^+e^-\to e^+e^-P$, see Fig.~\ref{fig:SL}. The common practice is to extract the TFF when one of the outgoing leptons is tagged and the other is not, that is, the single-tag method. The tagged lepton emits a highly off-shell photon with the momentum transfer $q_1^2\equiv -Q^2$ and is detected, while the other, untagged, is scattered at a small angle and its momentum transfer $q_2^2$ is almost zero, i.e., $F_{P\gamma^*\gamma}(Q^2)\equiv F_{P\gamma^*\gamma^*}(-Q^2,0)$.

Experimental information on its doubly virtuality can be obtained through the double-tag method which demands the tagging of both leptons in the final-state. Due to the decrease of the cross section for both virtual photons, and the difficulties of the detection of all the particles entering into the process, data for  the doubly virtuality are not yet available for the lowest pseudoscalars. 

Theoretically, the limits $Q^2=0$ and $Q^2\rightarrow\infty$ are well known in terms of the axial anomaly in the chiral-limit of QCD \cite{Adler:1969gk} and perturbative QCD \cite{Lepage:1980fj}, respectively. The TFF is then calculated as a convolution of a perturbative hard-scattering amplitude and a gauge-invariant meson distribution amplitude which incorporates the non-perturbative dynamics of QCD~\cite{Lepage:1980fj}. At that point, a model needs to be used either for the distribution amplitude or for the TFF itself~\cite{Lepage:1980fj}. The discrepancy among different approaches reflects the model-dependency of that procedure. On top, when the desired object should be parameterized for the whole space-like range, and available experimental data are going to be used, the asymptotic scale of QCD should be fixed and a contact with the low-energy description performed. Theoretical efforts have not yet succeeded on performing such an endeavor, even worse for ascribing a  theoretical errors to the procedure. A different strategy might be, then, desirable.

Let us remark that form factors are not interesting by themselves as represent the knowledge of QCD in a nutshell, but also for their important role on precision calculations of low-energy Standard Model observables such as the anomalous magnetic moment of the muon, the pseudoscalar decays into lepton pairs, the radius of the proton, etc, where deviations between the theoretical predictions and the experimental measure point towards an indirect search of New Physics phenomena.

The experimental information on the TFFs together with the theoretical knowledge on their kinematic limits yield the opportunity for a nice synergy between experiment and theory in a simple, easy, systematic, and user-friendly method. This synergy is the purpose of our work.

In this respect, our proposal can be summarized taking the following considerations:

\begin{itemize}
\item We want a \underline{method}, not a \textit{model}. Our proposal is actually a TOOLKIT.
\item It is a method based on the \underline{analyticity} and \underline{unitary} of the TFF, important ingredients when heading towards errors at the $10\%$ level.
\item Should be a method as \underline{simple} as possible, maximally transparent.
\item If possible, the method should not use any assumption, only \underline{approaches}. That means, a method improvable without \textit{ad hoc} statements.
\item We shall provide a \underline{systematic} method in two different senses: easy to update whenever new experimental data or new theoretical calculations are available; capable of provide a purely theoretical error from the approaches performed.
\item Finally, should be \underline{predictive} and checkable.
\end{itemize}

The answer to this catalogue of wishes can be found within the Theory of Pad\'e approximants. The connexion with the mathematical problem is given by the well-defined \textit{general rational Hermite interpolation problem}. This problem corresponds with the situation where a function should be approximated but previous information about it is scarce and spread over certain information on a given set of points together with a set of derivatives. Our goal is to make a contact with this problem from our needs and provide a model-independent and data-driven parameterization of the TFFs useful for the problems we have at the precision low-energy frontier of the Standard Model.

\subsection{Why Pad\'e Theory?}

In front of other more sophisticated models and methods, it may seem curious to appeal now to an old tool such as Pad\'e Theory to perform precise calculations within the low-energy frontier of the Standard Model. In this respect, two considerations must be taken into account. First, Pad\'e Theory is an active field of research in applied mathematics. Few of the results considered here have been the subject of research in the last years only. Second, the corpus of Pad\'e Theory defines precisely the problem we have at hand (as we already announced, the general rational Hermite interpolation problem) and provides with a solution in the form of convergence theorems and tools.

From a theoretical point of view, one should notice that for precise Standard Model calculations at low energies considered here, what is needed is not directly the TFF but an integral over the TFF with a particular weight. (The exception to this point discussed in this talk is the extraction of the $\eta-\eta'$ mixing parameters from the corresponding TFFs.) 

Another interesting assertion is the fact that parameterizations such as the Vector Meson Dominance, Lowest Meson Dominance (and extensions), models from holographic QCD~\cite{CapriProceedings} are already a certain kind of Pad\'e approximant (the so-called Pad\'e-Type approxiamnts~\cite{Masjuan:2007ay} where the poles of the Pad\'e are given in advanced). These parameterizations should take advantage of the theory of Pad\'e approximants if a robust calculation is to be claimed. For example, the uncertainties due to the truncation of the PA sequence which are never accounted for when using these parameterizations, do not need to be small to be neglected~\cite{Masjuan:2014rea}\footnote{In Ref.~\cite{Masjuan:2014rea} we showed how the correct treatment of these parameterizations \textit{\'a la Pad\'e} may yield results up to $20\%$ higher, an effect that goes beyond any attempt of a theoretical quoted error so far.}. On top, it is proven that Pad\'e Type approximants converge slower than PAs, specially when considering integrals up to infinity (cf. second Ref. in~\cite{Baker}). In this respect, heading towards $10\%$ errors for the pseudoscalar contributions to the HLBL (the typical size of the foreseen experimental uncertainty in the new experiment at FermiLab), all these considerations are a must.

The PAs feature an interesting relation with dispersion relations. The state-of-the-art dispersive formalism (DR) accounting for the HLBL (see Ref.~\cite{Massimo} together with the discussion in Ref.~\cite{Vainstein}) has to face the lack of high-energy constraints coming from QCD (cf. as well~\cite{CapriProceedings}). In this respect, PAs can provide a natural tool to extend the DR results and interpolate up to the high-energy region, including the well-known pQCD behavior into the game as it should. Actually, the PA themselves can be used by the DR as a parameterization where to impose the dispersive constraints, obtain the PA coefficients, and then extrapolate far away in the space-like region. From another point of view, the complementarity between PA and DR may come from the interplay between TL and SL regions. In this case, DR after accounting for the TL data (a region not accessible by the PA built from the SL), can provide the LECs to be used for reconstructing the PA in the SL. In this situation, PA would represent an extension of the DR into the SL. In an interactive procedure, SL experimental data can be used, as well, to constrain the PA parameters for later on being used back in the DR and so on.

Last but not least, as a user-friendly and simple tool, PA can be used in the analysis of experimental results as a corpus of fitting functions. In this situation, instead of using a VMD to fit the experimental data, the highest PA would yield the most accurate theoretical result including systematic errors without the need to use a particular model.

\section{Transition form factors from rational approximants}

The toolkit we have introduced is a powerful method to deal with functions of one and two variables for which experimental data points are known and theoretical constraints available. The method is, of course, general. We developed it focusing on the need to properly introduce experimental data in a systematic manner in observables that manifest a discrepancy with the Standard Model calculation. For the HLBL and pseudoscalar decays into lepton pairs, the object are meson TFFs. In this section we collect our main results for $\pi^0, \eta$ and $\eta'$.

We proposed in Refs.~\cite{Masjuan:2012wy,EscribanoMasjuan} to use a sequence of PA $P^N_M(Q^2)=\frac{{\cal P}_N(Q^2)}{{\cal P}_M(Q^2)}$ ~\cite{Baker}, to fit the space-like data~\cite{Aubert:2009mc,SL,Uehara:2012ag,Acciarri:1997yx}. Since PAs  are constructed from the Taylor expansion of the $F_{P\gamma^*\gamma}(Q^2)$, from the fits we can obtain the derivatives of the $F_{P\gamma^*\gamma}(Q^2)$ at the origin of energies in a simple, systematic and model-independent way~\cite{Masjuan:2012wy,EscribanoMasjuan}. The idea behind is that the TFF can be expanded as:
\begin{equation}\label{Taylor}
F_{P\gamma*\gamma} (Q^2,0)=a_0^P\big(1+b_{P} \frac{Q^2}{m_{P}^2}+c_{P} \frac{Q^4}{m_{P}^4}+ \dots\big)\, ,
\end{equation}
\noindent
where $a_0^P$ is related to the decay of the pseudoscalar into two photons through  
\begin{equation}\nonumber
|F_{P\gamma\gamma}(0)|^{2}=\frac{64\pi}{(4\pi\alpha)^{2}}\frac{\Gamma(P\rightarrow\gamma\gamma)}{m_{m_{P}}^{3}}.
\label{Pgammagamma0}
\end{equation}
$b_P$ and $c_P$ in Eq.~(\ref{Taylor}) are the slope and curvature of the TFF respectively. The low-energy parameters are fundamental quantities for constructing the PAs and any hadronic model to be used to evaluate hadronic contributions to the observables we are discussing.

\begin{figure}
\centering
\includegraphics[width=7.0cm,clip]{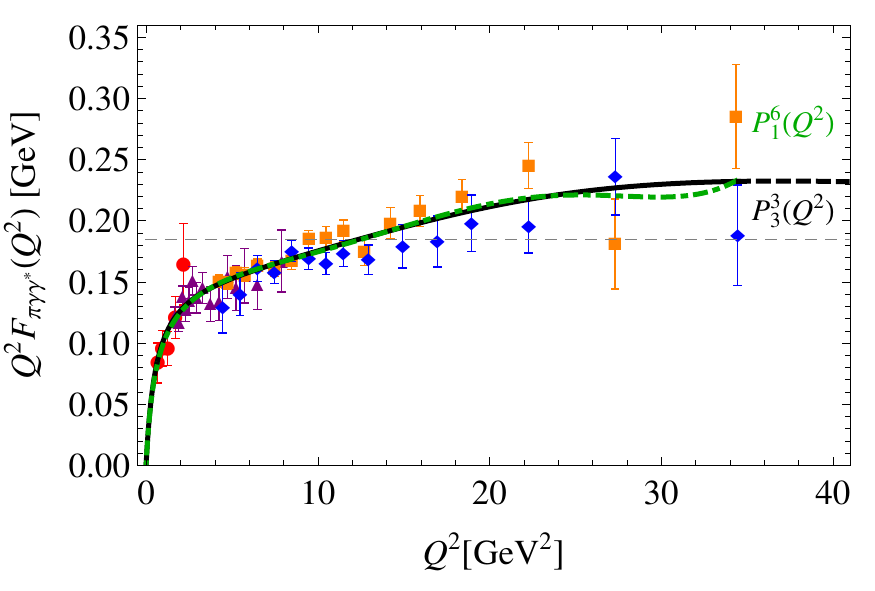}
\includegraphics[width=7.0cm,clip]{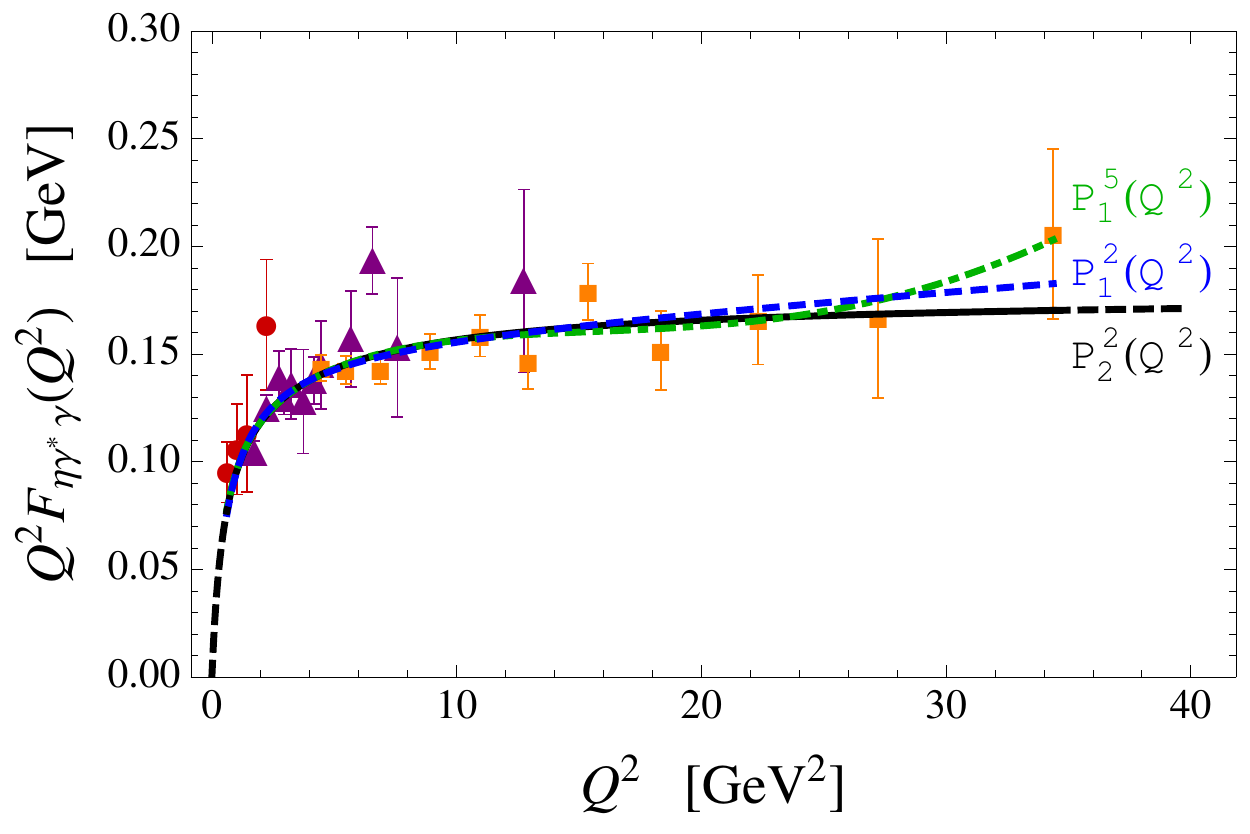}
\includegraphics[width=7.0cm,clip]{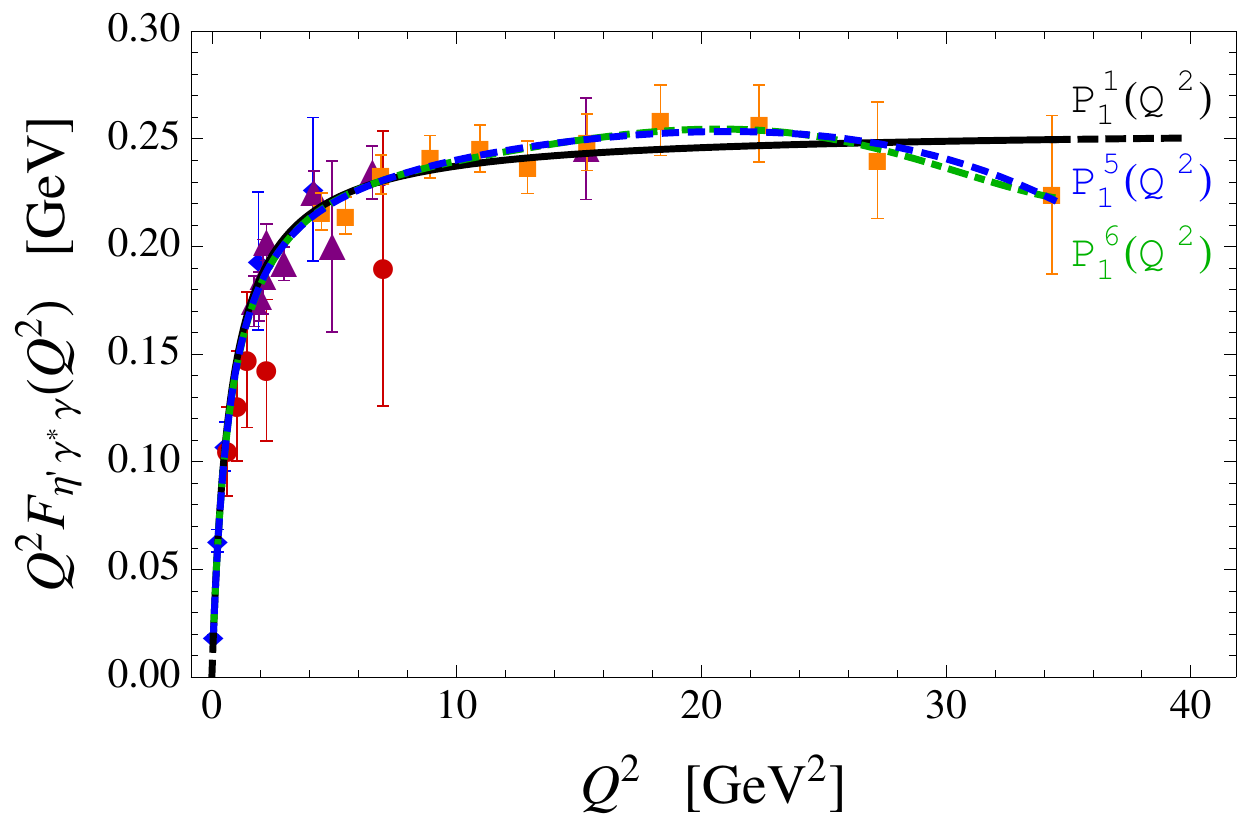}
\caption{$\pi^0$ (upper panel), $\eta$ (middle panel), and $\eta^\prime$ (lower panel) TFFs. 
Green-dot-dashed lines show our best $P^L_{1}(Q^2)$ fit, and black-solid lines show our best $P^N_{N}(Q^2)$ fit. Black-dashed lines display the extrapolation of the $P^N_{N}(Q^2)$ at $Q^2=0$ and $Q^2\to\infty$. Experimental data are from CELLO (red circles), CLEO (purple triangles), and {\textit{BABAR}} (orange squares) Colls.~\cite{SL}. The $\pi^0$ figure contains also data from BELLE (blue diamonds)~\cite{Uehara:2012ag}; and the $\eta^\prime$ figure data from L3 (blue diamonds) \cite{Acciarri:1997yx}. Figures form Refs.~\cite{Masjuan:2012wy,EscribanoMasjuan}.}
\label{fig1}       
\end{figure}

Since the analytic properties of TFFs are not known in general although the time-like region at low energies show the well-known unitary $\pi\pi$ cut, the kind of PA sequence to be used is not determined in advance. We consider two different sequences and the comparison among them should reassess our results. The first one is a $P^L_1(Q^2)$ sequence inspired by the success of the simple vector meson dominance ansatz~\cite{VFF}, and the second one is a $P^N_N(Q^2)$ sequence which satisfy the pQCD constrains $Q^2 F_{P\gamma \gamma^*}(Q^2) \sim \textrm{constant} $. After combining both sequences' results, slope and curvature from Refs.~\cite{Masjuan:2012wy,EscribanoMasjuan,Escribano:2015nra,Escribano:2015yup} are shown in Table~\ref{tab1}, where $\lim_{Q^2\to\infty }Q^2F_{P\gamma^*\gamma}(Q^2) $ from the $P^N_N(Q^2)$ is also shown. The results are shown in Fig.~\ref{fig1}.

\begin{table}
\centering
\caption{$\pi^0,\eta$, and $\eta^\prime$ slope $b_P$, curvature $c_P$, and asymptotic limit ($Q^2 \to \infty$) from Refs.~\cite{Masjuan:2012wy,EscribanoMasjuan,Escribano:2015nra}.}
\label{tab1}       
\begin{tabular}{cccc}
\hline
 & $b_P$ & $c_P$ & $\lim_{Q^2\to\infty }Q^2F_{P\gamma^*\gamma}(Q^2) $    \\\hline
$\pi^0$ & $0.0324(22)$ & $1.06(27)\cdot 10^{-3}$ & $2 f_{\pi} $   \\
$\eta$ & $0.60(7)$ & $0.37(12)$ & $0.160(24) \textrm{GeV}$ \\ 
$\eta^\prime$ & $1.30(17)$ & $1.72(58)$ & $0.255(4) \textrm{GeV}$  \\ \hline
\end{tabular}
\end{table}

\section{Applications of the transition form factors}
\subsection{Hadronic light-by-light scattering contribution to the muon $(g-2)$}

\begin{table*}
\caption{Preliminary results for the pseudoscalar pole contributions to  \label{tab:g2res} $a^{HLBL;\pi^0\eta,\eta'}_{\mu}$ for different values in the chosen $a_{1,1}$ range discussed in Ref.~\cite{Sanchez-Puertas:2015yxa}.}
\begin{center}
  \begin{tabular}[b]{lllll} 
  \hline
         Units of $10^{-10}$      & $\pi^0$ & $\eta$ & $\eta'$  &  Total \\ 
  \hline
   {\small{$a_{1,1} = 2b_{P}^2 \ \ [OPE]$}} \hspace{0.2cm}   & $6.64(33)$ & $1.69(6)$ & $1.61(21)$  &  $9.94(40)_{stat}(50)_{sys}$ \\
   {\small{$a_{1,1} = b_{P}^2$ \ \ [Fact]}}    & $5.53(27)$ & $1.30(5)$ & $1.21(12)$  &  $8.04(30)_{stat}(40)_{sys}$ \\
   {\small{$a_{1,1} = 0$}}   & $5.10(23)$ & $1.16(7)$ & $1.07(15)$  &  $7.33(28)_{stat}(37)_{sys}$ \\ 
  \hline
  \end{tabular}
  \end{center}
  \label{g2values}
\end{table*}

The HLBL cannot be directly related to any measurable cross section and requires the knowledge of QCD contributions at all energy scales. Since this is not known yet, one needs to rely on hadronic models to compute it. Such models introduce systematic errors which are difficult to quantify. The large-$N_c$ limit of QCD~\cite{largeNc} provides a very useful framework to approach this problem and has been used to perform what nowadays are the reference numbers for the HLBL~\cite{CapriProceedings}. This, however, has a shortcoming. Calculations carried out in the large-$N_c$ limit demand an infinite set of resonances. As such sum is not known in practice, one ends up truncating the spectral function in a resonance saturation scheme, the so-called Minimal Hadronic Approximation \cite{Peris:1998nj}. The resonance masses used in each calculation are then taken as the physical ones from PDG \cite{Agashe:2014kda} instead of the corresponding (but unknown) masses in the large-$N_c$ limit. Both problems might lead to large systematic errors not included so far \cite{Masjuan:2007ay,Masjuan:2012wy,Masjuan:2012gc}.

It was pointed out in Ref.~\cite{Masjuan:2007ay} that, in the large-$N_c$ framework, the Minimal Hadronic Approximation can be understood from the mathematical theory of PAs to meromorphic functions. Obeying the PA rules, one can compute the desired quantities in a model-independent way, within the large-$N_c$, and even be able to ascribe a systematic error to the approach~\cite{Masjuan:2009wy}.

Our proposal is, then, to use a sequence of PAs built up from the low-energy expansion obtained in the previous section. The form factor needed in this calculation is the one of double virtuality. For this reason we extend the PA method to the Canterbury approximants (CA)~\cite{Masjuan:2015lca}. Now, the Taylor expansion of the doubly-virtual TFF reads:
\begin{equation}\nonumber
F_{P\gamma*\gamma*} (Q_1^2,Q_2^2)=a_0^P\left(1+b_{P} \frac{Q_1^2+ Q_2^2}{m_{P}^2}+ a_{1,1} \frac{Q_1^2 Q_2^2}{m_{P}^4} + \dots\right)\, ,
\end{equation}
\noindent
where $a_{1,1}$ corresponds to the slope of double virtuality and we have made use of the Bose symmetry to simplify our expression. Having experimental data in some energy region, even without high precision ($30\%$ or even $50\%$ statistical error), one can attempt the TFF reconstruction in a systematical way by a sequence of doubly virtual approximants fitted to such data. CAs can accommodate the high-energy constraints from QCD as well~\cite{Masjuan:2015lca}:

\begin{equation}\label{Chisholm}
C^0_1(Q_1^2,Q_2^2)=\frac{a_0}{1+\frac{b_P}{m_P^2}(Q_1^2+Q_2^2)+\frac{2b_p^2-a_{1,1}}{m_P^4}Q_1^2Q_2^2}\, .
\end{equation}
\noindent
Knowing the Taylor expansion of the $F(Q_1^2,Q_2^2)$, Eq.~(\ref{Chisholm}) would be unique: $a_0$ is determined from the $\Gamma (\pi^0 \to \gamma\gamma)$ as before, $b_p$ is the slope of the single virtual TFF, and $a_{1,1}$ corresponds to the doubly-virtual slope.  

Experimental data for $F(Q_1^2,Q_2^2)$ is not available yet and we cannot extract $a_{1,1}$ from them. The OPE tells us that $\lim_{Q^2 \to \infty} F(Q^2,Q^2) \sim Q^{-2}$ and implies $a_{1,1}=2b_p^2$. On the other hand, at low energies, the factorization approach $F_{P\gamma^*\gamma^*}(Q_1^2,Q_2^2) = F_{P\gamma^*\gamma}(Q_1^2,0) \times F_{P\gamma \gamma^*}(0,Q_2^2)$ works well as it is already seen in the construction of the approximant~(\ref{Chisholm}) where the parameter $a_{1,1}$ enters as a correction to the $2b_p^2$. Taking the factorization into account, then $a_{1,1}=b_p^2$. To be on a conservative side, we consider for the numerical analysis $b_p^2 \leq a_{1,1} \leq 2 b_p^2$. This range should effectively take into account our ignorance on $a_{1,1}$. As soon as experimental data will be available, this range will be immediately shrunk. Our results for the pseudoscalar pole contribution to the HLBL for $\pi^0, \eta$ and $\eta'$ are collected in Table~\ref{g2values}. Let us remark that this is the most robust calculation of the $\eta$ and $\eta'$ contribution to the HLBL up to now. The range $(7.33 \div 9.94)\times 10^{-10}$ is as large as the pion loop contribution which is subleading in $1/N_c$. There is a large effort to estimate the pion loop at the $10\%$ precision using DRs~\cite{Massimo}. We notice, however, that this efforts will be in vain if the range we quote in Table~\ref{g2values}, last column, cannot be substantially reduced.  

On top, the TFF is considered to be off-shell. To match the large momentum behavior with short-distance constraints from QCD, calculable using the OPE, we consider the relations obtained in Ref.~\cite{CapriProceedings}. In practice this will amount to extend the CAs we are using to match the set of high-energy OPE constraints. These results are still preliminary and will be given elsewhere~\cite{inprep}. We anticipate, however, an increase of the results in Table~\ref{g2values}, last column, by about $20\%$, again of the order of the pion loop contribution.

The HLBL is dominate for $Q^2$ ranging from $0$ to $2$ GeV$^2$, in particular above around $0.5$ GeV$^2$. Therefore a good description of TFF in such region is very important. Such data are not yet available, but any model should reproduce the available one. That is why in~\cite{Masjuan:2012wy,EscribanoMasjuan}, in contrast to other approaches, we did not used data directly but the low-energy parameters of the Taylor expansion for the TFF and reconstructed it \textit{via} PAs. The LECs certainly know about all the data at all energies and as such incorporates all our experimental knowledge at once. This procedure implies a model-independent result together with a well-defined way to ascribe a systematic error. In other words, it is the first procedure that can be considered an \emph{approximation}, in contrast to the \emph{assumptions} considered in other approaches. 

Let us emphasize the role of experimental data by using the LECs in Table~\ref{tab1} together with the $\pi^0 \to \gamma \gamma$ to match the free parameters of the LMD+V model introduced in~\cite{Knecht:2001qf}. We would obtain $a_{\mu}^{\mathrm{HLBL},\pi^0}=7.5\times 10^{-10}$ which contrasts with the original $a_{\mu}^{\mathrm{HLBL},\pi^0}=6.3\times 10^{-10}$~\cite{Knecht:2001qf}. The impact of the new experimental data is then clear, inducing a $20\%$ effect. On top, since the LMD+V is not a PA but a well-educated model, it is difficult to ascribe a systematic error due to the large-$N_c$ approach. PAs in turn, already account for such corrections which in Refs.~\cite{Masjuan:2012wy,EscribanoMasjuan} were found to be of the order of $5\% - 10\%$ less dramatic than the naive $30\%$ from the $N_c$ counting, but still required to be taken into account.

\subsection{pseudoscalar decays into lepton paris}

\begin{figure}
\centering
\includegraphics[width=0.35\textwidth]{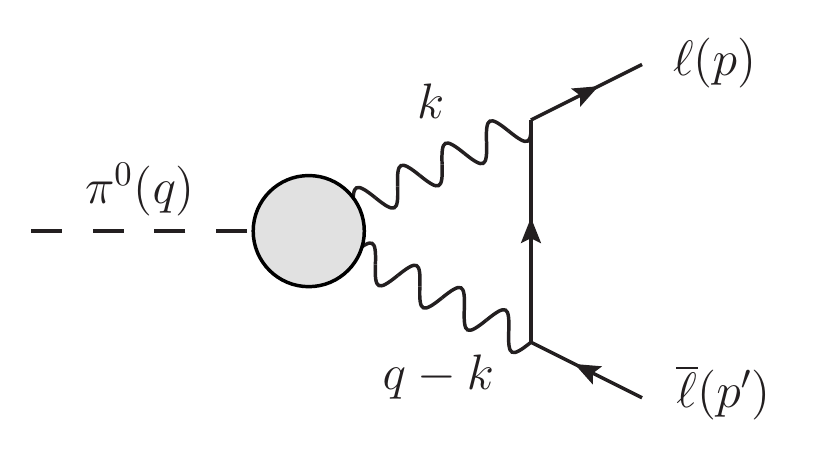}
\caption{\small{Feynman Diagram for $\pi^0 \to e^+e^-$ process.}}
\label{Pi0ee}
\end{figure}

Pseudoscalar decays into lepton pairs provide a unique environment for testing our knowledge of QCD. As such decays are driven by a loop process, encode, at once, low and high energies. For the $\pi^0$ decay, the process (neglecting electroweak corrections) proceeds in two steps as shown in Fig.~\ref{Pi0ee}.  
The loop does not diverge due to the presence of the pseudoscalar transition form factor on the $\pi^0 \to \gamma^* \gamma^*$ anomalous vertex~\cite{Adler:1969gk}, the $F_{P\gamma^*\gamma^*}(k^2,(q-k)^2)$ with $k^2,(q-k)^2$ space-like photon virtualities. For $\eta$ and $\eta'$ intermediate $\pi\pi$ cuts must be taken into account as well~\cite{Sanchez-Puertas:2015yba}.

The branching ratio (BR) for this decay may be expressed in terms of the two photon decay width as 
\begin{equation}\nonumber
\frac{\textrm{BR}(P\rightarrow\overline{\ell}\ell)}{\textrm{BR}(P\rightarrow\gamma\gamma)} = 2 \left( \frac{\alpha m_{\ell}}{\pi m_P}  \right)^2 \beta_{\ell} \left\vert\mathcal{A}(m_P^2)\right\vert^2,
\end{equation}
where $\mathcal{A}(q^2)$ represents the loop amplitude (see Ref.~\cite{Masjuan:2015lca} and references therein for details)
\begin{equation}\nonumber
\label{eq:loop}
\mathcal{A}(q^2) =  \frac{2i}{\pi^2q^2} \int d^4k \frac{[ q^2k^2 - (q\cdot k)^2]\tilde{F}_{P\gamma^*\gamma^*}(k^2,(q-k)^2)}{ k^2(q-k)^2((p-k)^2-m_{\ell}^2)}, 
\end{equation}
which is unknown as far as the normalized TFF $(\tilde{F}_{P\gamma\gamma}(0,0)=1)$ is unspecified. The role of the doubly virtual TFF is actually rather important as the 
given integral, similarly to the HLBL case, is UV-divergent. Remarkably, for the $\pi^0$ case, it is possible to go further without a single clue on the TFF. Being the 
lightest hadron, it is not possible to find any additional intermediate hadronic 
state which may be on-shell, and contribute therefore to the imaginary part. Consequently, its imaginary part is solely given by the intermediate $\gamma\gamma$ state, the well-known unitary bound discussed by Drell~\cite{Drell}, which gives
\begin{equation}
   \textrm{Im}(\mathcal{A}(m_{\pi^0}^2)) =  \frac{\pi}{2\beta_{\ell}}\ln\left( \frac{1-\beta_{\ell}}{1+\beta_{\ell}} \right) \quad (\beta_{\ell}=\sqrt{1-4m_{\ell}^2/m_{\pi^0}^2}),
\end{equation}

\begin{table*}
\centering
\begin{tabular}{cccc} 
\hline
Process & BR(th)  & BR(exp) \\ 
\hline
$\pi^0\rightarrow e^+e^-$      &  $(6.20\div 6.35)(5)\times10^{-8}$      & $7.48(38) \times 10^{-8}$~\cite{Abouzaid:2006kk}  \\ 
$\eta\rightarrow e^+e^-$        &  $(5.31 \div 5.44)(5)\times10^{-9}$     & $\leq 2.3\times10^{-6}$ ~\cite{Agakishiev:2013fwl} \\ 
$\eta\rightarrow \mu^+\mu^-$&  $(4.72 \div 4.52)(8)\times10^{-6}$    & $5.8(8)\times10^{-6}$ ~\cite{Abegg:1994wx}   \\ 
$\eta'\rightarrow e^+e^-$        & $(1.82 \div 1.86)(19)\times10^{-10}$ & $\leq 5.6\times10^{-9}$ ~\cite{Aulchenko:2014vkn,Akhmetshin:2014hxv} \\ 
$\eta'\rightarrow \mu^+\mu^-$ & $(1.37 \div 1.49)(33)\times10^{-7}$  & ---  \\ 
\hline
\end{tabular}
\caption{Preliminary results for the range $a_{1,1}\in(b_P^2\div 2b_P^2)$, i.e. $(OPE\div Fact)$.  The errors refers to the statistical error for BR$(P\rightarrow\gamma\gamma)$, the error from $b_P$ and the systematic, combined in quadrature.}
\label{tab:mainres}
\end{table*}

Due to the presence of the photon propagators, the kernel of the integral is peaked at around the electron mass. Then, one can expand that kernel in terms of $m_e/m_P$ but also $m_e/\Lambda$, being $\Lambda$ the cut off of the loop integral, or the hadronic scale driven by the TFF. \cite{Dorokhov:2007bd} resummed such power corrections and found them negligible~\cite{Dorokhov:2009xs}. Then, using a Vector Meson Dominance for the TFF they 
found $\mathrm{BR}(\pi^0\to e^+e^-)=(6.23 \pm 0.09)\times 10^{-8}$~\cite{Dorokhov:2009xs}, $3\sigma$ off the KTeV result.

Radiative corrections were recently reconsidered in~\cite{Vasko:2011pi} and yield the new KTeV result as $\mathrm{BR}_{\mathrm{"KTeV"}}(\pi^0\to e^+e^-)=(6.87 \pm 0.36)\times 10^{-8}$.

In~\cite{Masjuan:2015lca}, we investigated the role of the TFF of doubly virtuality using the CAs. Beyond considering the eventual effects of different New Physics scenarios (which are negligible), we consider for the numerical analysis that $0 \leq a_{1,1} \leq 2 b_P^2$, and obtained the new Standard Model value. Extending our calculations to the $\eta$ and $\eta'$ decays now including as well the $\mu \mu$ mode, the preliminary results are found in Table~\ref{tab:mainres}. 

The error comes from $P\to \gamma \gamma$ and $b_\pi$ together with the evaluation of the systematic error from our approximation~\cite{Masjuan:2015lca}, and the two main numbers from the ranging of $a_{1,1}$. To shrink the window here provided, experimental data would then be very welcome. For the $\pi^0$, this final number represents still a deviation of the measured BR between $2.6 - 1.4 \sigma$.

Forcing our approximant~(\ref{Chisholm}) to reproduce the KTeV result and then used for the $\pi^0$ contribution to the HLBL~\cite{CapriProceedings}, we obtain $a_{\mu}^{HLBL;\pi^0}=2.9 \times 10^{-10}$, a deviation of $2- 3 \times 10^{-10}$ with respect to the standard result. Taking into account that the global theoretical SM error for the muon $(g-2)$ is $6 \times 10^{-10}$~\cite{CapriProceedings}, the role of the $\pi^0 \to e^+e^-$ is certainly remarkable, and never been considered so far. Similar effect is also found for the $\eta \to \mu^+ \mu^-$ decay, indicating that the current precision of the SM error on the muon $(g-2)$ may be underestimated.

\subsection{$\eta-\eta^\prime$ mixing parameters}

The physical $\eta$ and $\eta'$ mesons are an admixture of the $SU(3)$ Lagrangian eigenstates~\cite{Leutwyler:1997yr}. Deriving the parameters governing the mixing is a challenging task. Usually, these are determined through the use of $\eta(')\rightarrow \gamma\gamma$ decays as well as vector radiative decays into $\eta(')$ together with $\Gamma(J/\Psi \to \eta^\prime \gamma)/\Gamma(J/\Psi \to \eta \gamma)$~\cite{Leutwyler:1997yr}. However, since pQCD predicts 
that the asymptotic limit of the TFF for the $\eta(')$ is essentially given in terms of these mixing 
parameters, we use our TFF parametrization to estimate the asymptotic limit and further constrain the mixing parameters with compatible results compared to standard (but more sophisticated) determinations~\cite{EscribanoMasjuan}. 

In this section we reanalyze~\cite{Escribano:2015yup} the $\eta-\eta'$ mixing as we did in Ref.~\cite{EscribanoMasjuan,Escribano:2015nra}. In these works, we took advantage of the flavor basis, where the $\eta$ and $\eta'$ pseudoscalar decay constants defined in terms of the axial current, $J_{5\mu}^a \equiv \overline{q}\gamma_{\mu}\gamma_5\frac{\lambda^a}{2} q$, as $\bra{0} J_{5\mu}^a \ket{P} \equiv ip_{\mu}F_P^a$ (where $a=q,s$ refers to light and strange quarks, resp.) can be expressed as
\begin{equation}\nonumber
(F_P^{a})\equiv
\begin{pmatrix}
F_{\eta}^{q}&F_{\eta}^{s}\\
F_{\eta^{\prime}}^{q}&F_{\eta^{\prime}}^{s}
\end{pmatrix}
=
\begin{pmatrix}
F_{q}\cos\phi_{q}&-F_{s}\sin\phi_{s}\\
F_{q}\sin\phi_{q}&F_{s}\cos\phi_{s}
\end{pmatrix},
\end{equation}
with $\lambda^q = \textrm{diag}(1,1,0)$ and $\lambda^s = \textrm{diag}(0,0,\sqrt{2})$. This basis has become popular since large-$N_c$ chiral perturbation theory NLO predictions 
yield~\cite{Escribano:2005qq,Feldmann:1998vh}
\begin{equation}\label{eqLambda1}
 F_qF_s \sin(\phi_q - \phi_s) = \frac{\sqrt{2}}{3} F_{\pi}^2\Lambda_1,
\end{equation}
where $F_{\pi}$ is the pion decay constant, and $\Lambda_1$ is an OZI-rule-violating parameter expected to be small. Assuming $\Lambda_1$ to be neglected, Eq.~(\ref{eqLambda1}) implies $\phi_q=\phi_s\equiv\phi$, an approximation 
which has been successful in phenomenological applications.

Large-$N_c$ ChPT predicts also that
\begin{equation}
 F_q^2 = F_{\pi}^2 + \frac{2}{3}F_{\pi}^2\Lambda_1\, .
\end{equation}

Here, phenomenological studies~\cite{EscribanoMasjuan,Escribano:2015nra,Escribano:2015yup,Escribano:2005qq,Feldmann:1998vh} do not support $\Lambda_1=0$ since they find $F_q>F_{\pi}$. Therefore, to be consistent we will consider the most general case $\phi_q\neq\phi_s$ and work in the so-called octet-singlet basis, where the decay constants are defined as 
\begin{equation}\nonumber
F_P^{80}\equiv
\begin{pmatrix}
F_{\eta}^{8}&F_{\eta}^{0}\\
F_{\eta^{\prime}}^{8}&F_{\eta^{\prime}}^{0}
\end{pmatrix}
=
\begin{pmatrix}
F_{8}\cos\theta_{8}&-F_{0}\sin\theta_{0}\\
F_{8}\sin\theta_{8}&F_{0}\cos\theta_{0}
\end{pmatrix}.
\end{equation}
In this basis, Large-$N_c \chi PT$ at NLO predicts~\cite{Escribano:2005qq,Feldmann:1998vh}
\begin{align}
 & F_8^2 = \frac{4F_K^2 - F_{\pi}^2}{3},  \qquad F_0^2 = \frac{2F_K^2 + F_{\pi}^2}{3} + F_{\pi}^2\Lambda_1, & \label{eq:chptF8F0} \\
 & (\theta_8 - \theta_0) = -\frac{4\sqrt{2}}{3}\left( \frac{F_K}{F_{\pi}} -1 \right)& \label{eq:chptangle} 
\end{align}
where $F_K$ is the kaon decay constant. To derive~\eqref{eq:chptF8F0} and \eqref{eq:chptangle} we made use of the relation between the low-energy constant $L_5$ appearing in the Large-$N_c \chi PT$ Lagrangian at NLO~\cite{Ecker:2010nc} and the ratio $F_K/F_\pi$. $F_0$ is renormalization group dependent. This is connected to the $J_{5\mu}^0$ anomalous dimension implying~\cite{Leutwyler:1997yr,Agaev:2014wna}
\begin{equation}\nonumber
\mu \frac{dF_0}{d\mu} = -N_F\left( \frac{\alpha_s(\mu)}{\pi} \right)^2 F_0,
\end{equation}
where $N_F$ is the number of active flavors. Solving this equation at order $\alpha_s$, 
\begin{align}
F_0(\mu) = & \ F_0(\mu_0)\left(1+\frac{2N_F}{\beta_0}\left( \frac{\alpha_s(\mu)}{\pi} - \frac{\alpha_s(\mu_0)}{\pi} \right) \right) \nonumber \equiv F_0(1+\delta),
\end{align}
with $\beta_0=11N_c/3 -2N_F/3$.
In the singlet-octet basis, the different limiting behaviors of the TFF, $F_{P\gamma\gamma}\equiv F_{P\gamma^*\gamma}(0)$, and $P_{\infty} \equiv \lim_{Q^2\to\infty}Q^2F_{P\gamma^*\gamma}(Q^2) $ 
take the simple form%
\begin{align}
 F_{\eta\gamma\gamma} = & \ \frac{1}{4\pi^2}\frac{\hat{c}_8 F_{\eta'}^0 - \hat{c}_0(1+\Lambda_3) F_{\eta'}^8}{F_{\eta'}^0 F_{\eta}^8 - F_{\eta'}^8 F_{\eta}^0}, \label{eq:Feta0} \\
 F_{\eta'\gamma\gamma} =  & \ \frac{1}{4\pi^2}\frac{-\hat{c}_8 F_{\eta}^0 + \hat{c}_0(1+\Lambda_3) F_{\eta}^8}{F_{\eta'}^0 F_{\eta}^8 - F_{\eta'}^8 F_{\eta}^0}, \label{eq:Fetap0}\\
 \eta_{\infty} =  & \ 2(\hat{c}_8 F_{\eta}^8 + \hat{c}_0(1+\delta) F_{\eta}^0), \label{eq:Infeta} \\
  \eta'_{\infty} = & \ 2(\hat{c}_8 F_{\eta'}^8 + \hat{c}_0(1+\delta) F_{\eta'}^0),\label{eq:Infetap}
\end{align}
where $\hat{c}_8 = 1/\sqrt{3}$ and $\hat{c}_0 = \sqrt{8/3}$ are charge factors and $\delta=-0.17$~\cite{Escribano:2015nra} accounts for the $F_0$ running up to 
$Q^2\rightarrow\infty$~\cite{Agaev:2014wna}. In addition, we have included the OZI-violating parameter $\Lambda_3$, which has been neglected in our previous studies since it enters at the same level as $\Lambda_1$.

The set of Eqs.~(\ref{eq:Feta0}-\ref{eq:Infetap}) form a system of 4 equations with 5 unknowns ($F_{\eta^{(')}}^{(8,0)}, \Lambda_3$). Then it could seem that, at least taking $\Lambda_3=0$, we may solve the system. However, as explained in~\cite{Escribano:2015nra}, such a system is not linear independent as there is the relation
\begin{equation}
\label{eq:deg}
\eta_{\infty}F_{\eta\gamma\gamma} + \eta'_{\infty}F_{\eta'\gamma\gamma} =   \frac{3}{2\pi^2}\left( 1 + \frac{8}{9}(\delta +\Lambda_3+\delta\Lambda_3)  \right)
\end{equation}
which is free of mixing parameters. Indeed, Eq.~\eqref{eq:deg} determines $\Lambda_3$ once its left-hand-side is (experimentally) known but, we still have to face the fact that our system is linear dependent. In order to overcome this problem, we notice that at NLO in large-$N_c \chi PT$, Eqs.~(\ref{eq:chptF8F0},\ref{eq:chptangle}), provides a clean prediction both for 
$F_8$ and $(\theta_8 - \theta_0)$ in terms of the well-known value for $F_K/F_{\pi}$~\cite{Agashe:2014kda}. Taking either $F_8$ or $(\theta_8-\theta_0)$ constraint, would add an additional equation to the previous system, which would provide a unique solution. Taking both, would lead to an overdetermined system, which in general has no solution. For this reason, we adopt a democratic procedure in which we perform a fit including both, $F_8$ and $(\theta_8 - \theta_0)$ constraints, together with Eqs.~(\ref{eq:Feta0}-\ref{eq:Infetap}). In addition, we include the theoretical uncertainties from Large-$N_c \chi PT$ predictions, Eqs.~(\ref{eq:chptF8F0},\ref{eq:chptangle}) by noticing that $F_K/F_{\pi}$ typically 
receives $5\%$ corrections from the NNLO~\cite{Ecker:2010nc}. Consequently, we add this error in quadrature on top of the one from~\cite{Agashe:2014kda} for our fitting procedure. 

The solutions are collected in Tab.~\ref{tab:81res}.
\begin{table*}
\centering
\begin{tabular}{ccccccc} \hline
$F_8/F_\pi$ & $F_0/F_\pi$ & $\theta_8$ & $\theta_0$ & $\Lambda_3$ & $R_{J/\Psi}$ & $\chi^2/DOF$ \\ \hline
$1.32(7)$ & $1.25(3)$ & $-22.8(1.1)$ & $-7.6(2.2)$ & $0.05(3)$ & 5.6(7)& $1.0$\\ \hline
\end{tabular}
\caption{Predictions for the mixing parameters in the singlet-octet basis. $F_\pi=92.2$MeV and $\theta_{8,0}$ in degrees.}
\label{tab:81res}
\end{table*}
The value for $\chi^2/DOF$ is excellent 
which indicates a good agreement with large-$N_c \ \chi PT$ but predicts non-negligible NNLO corrections accounted here as a $5\%$. Without this $5\%$, the $\chi^2/DOF$ would grow up to $1.5$. In addition we can use Eqs.~(\ref{eq:chptF8F0},\ref{eq:chptangle}), to predict the value for the OZI-rule--violating parameter $\Lambda_1=0.21(5)$.

\begin{figure*}[htb]
\begin{center}
	\includegraphics[width=0.43\textwidth]{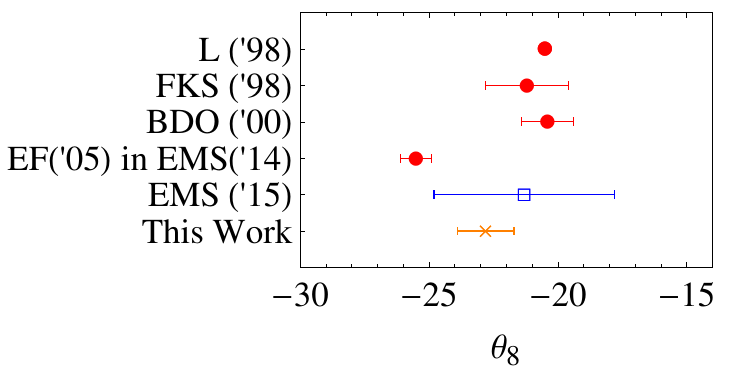}
	\hspace{-3.4cm}\includegraphics[width=0.43\textwidth]{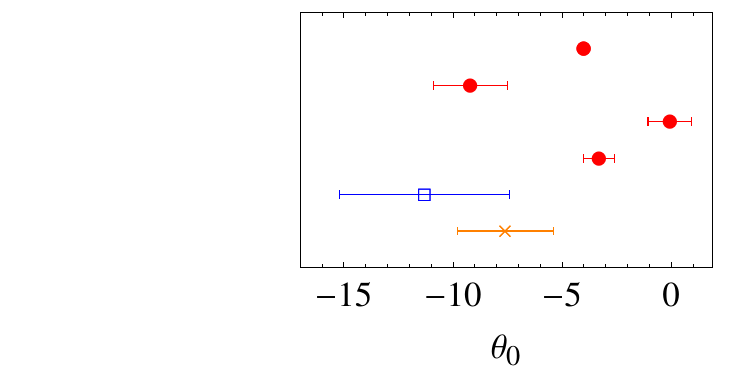}\\
	\includegraphics[width=0.43\textwidth]{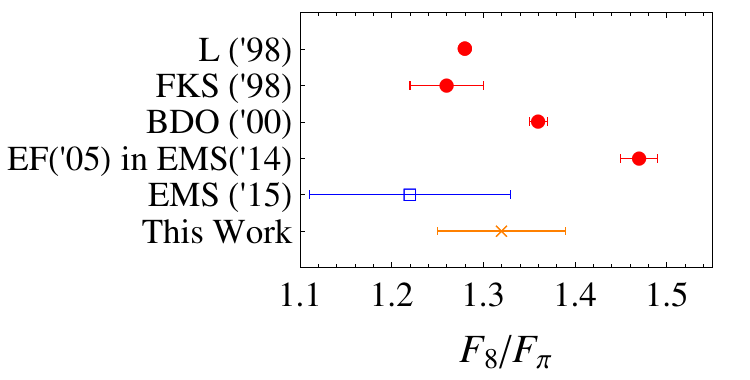}
	\hspace{-3.4cm}\includegraphics[width=0.43\textwidth]{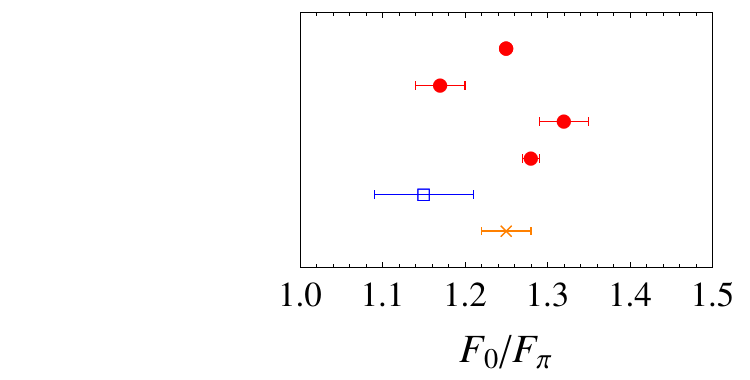}
\caption{$\eta-\eta^\prime$ mixing parameters in the octet-singlet basis from L~\cite{Leutwyler:1997yr}, FKS~\cite{Feldmann:1998vh}, BDO~\cite{Benayoun:1999au}, EF~\cite{Escribano:2005qq}, EMS(14)~\cite{EscribanoMasjuan}, EMS(15)~\cite{Escribano:2015nra}.}
\label{fig:mixing80}
\end{center}
\end{figure*}

In Fig.~\ref{fig:mixing80} we collect our results (orange crosses)~\cite{Escribano:2015yup} and compare them to different predictions in the literature~\cite{Leutwyler:1997yr,Feldmann:1998vh,Benayoun:1999au,Escribano:2005qq} (red dots), together with our previous results~\cite{Escribano:2015nra} in blue-empty squares. We see that the main difference with respect to our previous work~\cite{Escribano:2015nra}, where we did not use the $\eta'$ TFF asymptotic value and assumed $\phi_q=\phi_s$, appears in $F_0$. This is to be expected as the inclusion of $\Lambda_{1}$ and $\Lambda_{3}$ affects the singlet part exclusively. In addition, we have 
reduced our errors thanks to the constraints from Large-$N_c \chi PT$ with respect to our previous work. Our prediction for $\Lambda_3$ may be compared with the one in~\cite{Benayoun:1999au}, $\Lambda_3=-0.03(2)$. 
Both of them point towards a small value for this parameter, though they differ in sign. We find that $\Lambda_3$ actually plays an important role not only in fulfilling the degeneracy condition, Eq.~\eqref{eq:deg}, but in the $\eta(\eta')\rightarrow\gamma\gamma$ decays as well. In addition, the $\Lambda_1$ term is rather important and affects specially the $\eta'$ results, where deviations of order $10\%$ appear if this is omitted. Finally, we stress that the use of the RG equation for $F_0$ is fundamental, 
increases $\eta_{\infty}$ and diminishes $\eta'_{\infty}$, bringing in agreement experiment and theory. We encourage then the future analysis to account for it.

In addition, we can predict the ratio $R_{\JP} \equiv BR(\JP \rightarrow\eta'\gamma)/BR(\JP \rightarrow\eta\gamma)$, which is given in terms of 
$\phi_q$ alone~\cite{Escribano:2005qq} as 

$R_{\JP}=5.6(7)$, just at $1.2\sigma$ from the experimental value $R_{\JP}=4.7(2)$~\cite{Agashe:2014kda}.

\section{Conclusions and Outlook}

Pseudoscalar meson transition form factors are a good laboratory to study the properties of pseudoscalar mesons. Their interest goes, however, much beyond the mesons themselves as they play a key role on precision calculations of Standard Model observables at low energies where hadronic contributions are the cornerstone of the error evaluation. We propose the method of Pad\'e approximants as a \underline{toolkit} to analyze them. The method is easy, systematic, user-friendly, and can be improved upon by including new data. Provides, as well, information about the underlying structure of the TFF and can be used to extrapolate experimental information to extract the low-energy parameters of the form factor together with their asymptotic limits. The former makes contact with the experimental measurement of the pseudoscalar decay into two photons as well as calculations of the slope and curvature, and the later provides insights into the perturbative QCD regime.

Once LEPs and asymptotic values are known, the TFFs allow us to study the $\eta-\eta'$ mixing in a framework consistent with chiral perturbation theory within the large-$N_c$ limit at NLO including OZI-rule-violating parameters. 

The most relevant feature of the method here described is their excellent performance as an interpolation tool. As such, it is the most compelling method to provide an accurate parameterization for the TFF in the whole space-like region. This is of primordial importance for the correct assessment, in a data-driven approach, of the pseudoscalar contribution to the HLBL~\cite{Masjuan:2014rea,Sanchez-Puertas:2015yxa}. But this does not stop here. Since the approximants can as well penetrate into the time-like region below the first resonance, precise experimental data can be easily incorporated. Beyond the knowledge the approximants bring with respect to the features of the TFFs at time like, with this extension pseudoscalar decays into lepton pairs can be studied. In this regards, our method provides the most accurate data-driven and model-independent result consistent not only with the well-known QCD features at high and low energies, but as well with a well with a mathematical theory. This excellent performance is here exemplified by calculating the pseudoscalar pole contributions to the HLBL  and pseudoscalar decays into leptons pairs (for $\pi^0,\eta$ and $\eta'$) resulting in the most updated and data-driven results for these quantities up to now.

\section*{Acknowledgments}

Work partially supported by the Deutsche Forschungsgemeinschaft DFG through the Collaborative Research Center ``The Low-Energy Frontier of the Standard Model" (SFB 1044). P. M. wants to thank the organizers of the FCCP2015 conference for encouragement and support.

\end{document}